\documentclass[12pt,a4paper]{iopart}

\usepackage{iopams}
\usepackage{setstack}
\usepackage{graphicx}
\usepackage{pict2e}
\usepackage{color}
\usepackage{hyperref}

\newcommand{\sgn}{\mathrm{sgn}}
\newcommand{\C}[2]{{#1 \choose #2}}
\newcommand{\Li}{\mathrm{Li}}
\renewcommand{\Re}{\mathrm{Re}}
\renewcommand{\Im}{\mathrm{Im}}
\newcommand{\openone}{{\bf1}}
\newcommand{\half}{\tfrac{1}{2}}
\renewcommand{\vec}[1]{\mathbf{#1}}

\newcommand{\affiliation}[1]{\address{#1}}
\renewcommand{\pacs}[1]{\noindent\textbf{PACS numbers:} #1}
\newcommand{\keywords}[1]{\noindent\textbf{Keywords:} #1}
\newcommand{\tfrac}[2]{\mbox{\small$\frac{#1}{#2}$}}
\renewcommand{\text}[1]{\mathrm{#1}}

\begin{document}

\title[Extrapolation methods and Bethe ansatz for ASEP]{Extrapolation methods and Bethe ansatz for the asymmetric exclusion process}
\author{Sylvain Prolhac}
\affiliation{Laboratoire de Physique Th\'eorique, IRSAMC, UPS, Universit\'e de Toulouse, CNRS, France}

\begin{abstract}
The one-dimensional asymmetric simple exclusion process (ASEP), where $N$ hard-core particles hop forward with rate $1$ and backward with rate $q<1$, is considered on a periodic lattice of $L$ site. Using KPZ universality and previous results for the totally asymmetric model $q=0$, precise conjectures are formulated for asymptotics at finite density $\rho=N/L$ of ASEP eigenstates close to the stationary state. The conjectures are checked with high precision using extrapolation methods on finite size Bethe ansatz numerics. For weak asymmetry $1-q\sim1/\sqrt{L}$, double extrapolation combined with an integer relation algorithm gives an exact expression for the spectral gap up to $10$-th order in the asymmetry.\\\\
\\\\
\end{abstract}

\maketitle

\begin{section}{Introduction}
\label{section introduction}
The one-dimensional asymmetric simple exclusion process (ASEP) \cite{D2007.1,GM2006.1} is a Markov process describing hard-core particles moving locally on a lattice, with the exclusion constraint that each site is either empty or contains a single particle. The particles hop of one site in the forward direction with rate $1$ and of one site in the backward direction with rate $q$, $0\leq q<1$, if the destination site is empty. We consider in this paper the model with $N$ particles on a periodic lattice of $L$ sites, in the limit $L,N\to\infty$ with fixed density $\rho=N/L$, $0<\rho<1$.

It is expected that the only influence of the asymmetry parameter $q$ on the large scale behaviour is a rescaling of time by a factor $1-q$ for fixed $q<1$. This was shown for the model defined on the infinite line $\mathbb{Z}$ \cite{J2000.1,TW2009.1} and for stationary fluctuations in the periodic model \cite{DL1998.1,LK1999.1}. More generally, ASEP belongs at large scale to KPZ \cite{KPZ1986.1} universality \cite{KK2010.1,SS2010.4,QS2015.1,HHT2015.1}, which also describes fluctuations in some classes of driven diffusive systems, interface growth models and directed polymers in random media. KPZ universality is characterized in one dimension by a dynamical exponent $z=3/2$. For a finite system of $L$ sites, it leads to different behaviours at short time and long time, with a crossover on the relaxation scale $t\sim L^{3/2}$. The short time behaviour corresponds to the model defined on the infinite line, where there is much evidence for KPZ universality \cite{F2010.1,C2011.1}, and where fluctuations are characterized by Tracy-Widom distributions from random matrix theory. The long time behaviour corresponds to the non-equilibrium steady state, and there is also reasonable evidence for the universality of KPZ fluctuations there, with the Derrida-Lebowitz stationary large deviation function appearing for several models \cite{DL1998.1,DA1999.1,LK1999.1,BD2000.1,PPH2003.1,GLMV2012.1}.

ASEP is an exactly solvable model. Its generator is the Hamiltonian of a twisted XXZ spin chain with anisotropy $\Delta=(q^{1/2}+q^{-1/2})/2>1$, which is diagonalizable using Bethe ansatz. The study of the large scale behaviour of ASEP requires asymptotics of the eigenvalues and the eigenvectors of the generator. Very explicit formulas were obtained recently \cite{P2016.1} for such asymptotics in the totally asymmetric model (TASEP) $q=0$, due to the special structure of the Bethe equations when $\Delta\to\infty$, leading to exact expressions for fluctuations on the relaxation scale. Another approach based on the propagator is also possible \cite{BL2016.1}. Under the assumption that KPZ universality holds in the crossover between the short and long time limits, the TASEP asymptotics should also be valid for ASEP with $0<q<1$. Precise conjectures are formulated in this paper for the eigenstates. They are checked numerically with high precision using powerful extrapolation methods applied to finite size Bethe ansatz numerics. The extrapolation methods also allow to probe the weakly asymmetric regime $1-q\sim1/\sqrt{L}$ corresponding to a crossover to equilibrium fluctuations, which we illustrate for the spectral gap.

In section \ref{section Bethe ansatz}, we summarize various finite size Bethe ansatz formulas for the eigenstates of ASEP, with the algebraic formulation of Bethe ansatz recalled in \ref{appendix algebraic Bethe ansatz}, and numerical schemes for solving the Bethe equations discussed in \ref{appendix numerics Bethe roots}. Conjectures for asymptotics of ASEP eigenstates and their consequences for current fluctuations are stated in section \ref{section asymptotics}. The asymptotics are checked numerically using extrapolation methods in section \ref{section numerics}, where the case of the spectral gap for weak asymmetry is also discussed.
\end{section}

\begin{section}{Bethe ansatz for ASEP}
\label{section Bethe ansatz}
In this section, we summarize known finite size Bethe ansatz formulas for the eigenstates of ASEP. Some of them were already known in the context of ASEP, others are translated from the literature on the XXZ spin chain.

\begin{subsection}{Master equation}
ASEP is a Markov process. The probability $P_{t}(\mathcal{C})$ to observe the system in a configuration $\mathcal{C}$ at time $t$ is given by a master equation. Then
\begin{equation}
|P_{t}\rangle=\rme^{tM}|P_{0}\rangle\;
\end{equation}
with $M$ the Markov matrix, $\langle\mathcal{C}|P_{t}\rangle=P_{t}(\mathcal{C})$, and $\langle\mathcal{C}_{0}|P_{0}\rangle$ the initial probabilities at time $0$. We consider the (time-integrated) current $Q_{0}$ between site $L$ and site $1$ up to time $t$, defined as the number of moves of particles from site $L$ to site $1$ minus the number of moves from site $1$ to site $L$. The generating function of $Q_{0}$ can be computed using a deformation $M(\gamma)$ of the Markov matrix, with $\gamma$ a fugacity conjugate to the current and $M(0)=M$. The deformed generator $M(\gamma)$ can be written $M(\gamma)=\sum_{i=1}^{L}M_{i,i+1}(\gamma)$ with local operators $M_{i,i+1}(\gamma)$ acting only on sites $i$ and $i+1$. Their matrix in the local basis $(1_{i}\,1_{i+1},\,1_{i}\,0_{i+1},\,0_{i}\,1_{i+1},\,0_{i}\,0_{i+1})$, with $1$ and $0$ denoting respectively occupied and empty sites, is
\begin{equation}
M_{i,i+1}(\gamma)=
\left(
\begin{array}{cccc}
 0 & 0 & 0 & 0\\
 0 & -1 & q\,\rme^{-\gamma} & 0\\
 0 & \rme^{\gamma} & -q & 0\\
 0 & 0 & 0 & 0
\end{array}
\right)_{\!\!i,i+1}\;.
\end{equation}
The generating function of the current is then equal to \cite{P2016.1}
\begin{equation}
\label{GFcurrent[M(gamma)]}
\langle\rme^{L\gamma Q_{0}}\rangle=\sum_{\mathcal{C}}\langle\mathcal{C}|\rme^{-\gamma S}\rme^{tM(\gamma)}\rme^{\gamma S}|P_{0}\rangle\;.
\end{equation}
The operator $S$, diagonal in configuration basis, is defined by $S|\mathcal{C}\rangle=\big(\sum_{j=1}^{N}x_{j}\big)|\mathcal{C}\rangle$. The $x_{j}$'s, $1\leq x_{j}\leq L$, are the positions of the particles counted from site $1$.
\end{subsection}

\begin{subsection}{Bethe ansatz}
The matrix $M(\gamma)$ can be diagonalized using Bethe ansatz, see \textit{e.g.} \cite{DL1998.1}. Each eigenstate is completely characterized by $N$ complex numbers $y_{j}$, $j=1,\ldots,N$, the Bethe roots, that satisfy a set of $N$ polynomial equations called the Bethe equations:
\begin{equation}
\label{Bethe equations}
\rme^{L\gamma}\Big(\frac{1-y_{j}}{1-qy_{j}}\Big)^{L}=-\prod_{k=1}^{N}\frac{y_{j}-qy_{k}}{qy_{j}-y_{k}}\;.
\end{equation}
The corresponding eigenvalue of $M(\gamma)$ is equal to
\begin{equation}
E=(1-q)\sum_{j=1}^{N}\Big(\frac{1}{1-y_{j}}-\frac{1}{1-qy_{j}}\Big)\;,
\end{equation}
and the eigenvalue of the translation operator is
\begin{equation}
\rme^{2\rmi\pi p/L}=\rme^{N\gamma}\prod_{j=1}^{N}\frac{1-y_{j}}{1-qy_{j}}\;
\end{equation}
with $p\in\mathbb{Z}$ the total momentum.

The right and left eigenvectors of $M(\gamma)$, defined from the algebraic Bethe ansatz in \ref{appendix algebraic Bethe ansatz}, can be written explicitly in coordinate form. For a configuration with positions $\vec{x}$ ordered as $1\leq x_{1}<\ldots<x_{N}\leq L$, one has
\begin{eqnarray}
\label{psiR coordinate}
&& \langle\vec{x}|\psi\rangle=(1-q)^{N}\rme^{-\frac{N(N+1)}{2}\gamma}\sum_{\sigma\in\mathcal{S}_{N}}\prod_{j=1}^{N}\Bigg(\frac{\rme^{\gamma x_{j}}y_{j}}{1-y_{j}}\bigg(\frac{1-y_{\sigma(j)}}{1-qy_{\sigma(j)}}\bigg)^{x_{j}}\Bigg)\\
&&\hspace{60mm} \times\prod_{j=1}^{N}\prod_{k=j+1}^{N}\frac{y_{\sigma(j)}-qy_{\sigma(k)}}{y_{\sigma(j)}-y_{\sigma(k)}}\;\nonumber\\
\label{psiL coordinate}
&& \langle\psi|\vec{x}\rangle=(1-q)^{N}\rme^{-\frac{N(N+1)}{2}\gamma}\sum_{\sigma\in\mathcal{S}_{N}}\prod_{j=1}^{N}\Bigg(\frac{\rme^{\gamma(1-x_{j})}}{1-y_{j}}\bigg(\frac{1-y_{\sigma(j)}}{1-qy_{\sigma(j)}}\bigg)^{1-x_{j}}\Bigg)\\
&&\hspace{60mm} \times\prod_{j=1}^{N}\prod_{k=j+1}^{N}\frac{qy_{\sigma(j)}-y_{\sigma(k)}}{y_{\sigma(j)}-y_{\sigma(k)}}\;,\nonumber
\end{eqnarray}
with $\mathcal{S}_{N}$ the set of $N!$ permutations of $\{1,2,\ldots,N\}$. Since $M(\gamma)$ is not Hermitian, the left and right eigenstates are different. They verify however $(\prod_{j=1}^{N}y_{j}^{-1})\langle\vec{x}|\psi\rangle=\langle\psi|\vec{\tilde{x}}\rangle$ with $\vec{\tilde{x}}$ related to $\vec{x}$ by space reversal, $\tilde{x}_{j}=L+1-x_{N+1-j}$, $j=1,\ldots,N$, since transposing $M(\gamma)$ is the same as reversing space.

Completeness of the Bethe ansatz is the hypothesis that the Bethe equations (\ref{Bethe equations}) have exactly $\Omega=\C{L}{N}$ acceptable solutions such that the corresponding eigenvectors form a complete basis of the space generated by configurations with $N$ particles. Completeness is widely believed to be true based on numerics for small systems, but is difficult to prove rigorously, see however \cite{LSA1995.1,LSA1997.1,BDS2015.1} for mathematical results for the exclusion process.
\end{subsection}

\begin{subsection}{Scalar product of Bethe eigenstates}
The Bethe eigenstates are not normalized. Their scalar product is given by the Gaudin determinant \cite{GMCW1981.1,K1982.1}
\begin{eqnarray}
\label{norm[y]}
&&\fl\hspace{15mm} \langle\psi|\psi\rangle=(-1)^{N}(1-q)^{N}\rme^{-N^{2}\gamma}\Bigg(\prod_{j=1}^{N}y_{j}\Bigg)\Bigg(\prod_{j=1}^{N}\prod_{k=j+1}^{N}\frac{(y_{j}-qy_{k})(qy_{j}-y_{k})}{(y_{j}-y_{k})^{2}}\Bigg)\nonumber\\
&&\fl\hspace{40mm} \times\det\Bigg(\partial_{y_{i}}\log\bigg(\Big(\frac{1-y_{j}}{1-qy_{j}}\Big)^{L}\prod_{k=1}^{N}\frac{qy_{j}-y_{k}}{y_{j}-qy_{k}}\bigg)\Bigg)_{i,j=1,\ldots,N}\;.
\end{eqnarray}
The derivative with respect to $y_{i}$ in the determinant has to be computed before setting the $y_{j}$'s equal to a solution of the Bethe equations (\ref{Bethe equations}). At $q=0$, the determinant can be calculated explicitly \cite{MSS2012.2}, which allows the asymptotic analysis for large $L$, $N$ \cite{P2015.2}.

The Gaudin determinant is a consequence of the Slavnov determinant \cite{S1989.1} for the scalar product between a Bethe eigenstate $\psi_{\vec{y}}$ with Bethe roots $y_{j}$ and a Bethe vector $\psi_{\vec{w}}$ with arbitrary parameters $w_{j}$ not solution of Bethe equations,
\begin{eqnarray}
\label{Slavnov}
&& \langle\psi_{\vec{y}}|\psi_{\vec{w}}\rangle=(-1)^{\frac{N(N-1)}{2}}\Big(\prod_{j=1}^{N}\prod_{k=1}^{N}(w_{j}-y_{k})\Big)\\
&&\hspace{30mm} \times\Big(\prod_{j=1}^{N}\prod_{k=j+1}^{N}\frac{1}{(w_{j}-w_{k})(y_{j}-y_{k})}\Big)\det\Big[\partial_{y_{j}}E(w_{k})\Big]_{j,k}\;.\nonumber
\end{eqnarray}
The quantity
\begin{equation}
\label{eigenvalue T}
E(w)=\rme^{-N\gamma}\prod_{j=1}^{N}\frac{w-qy_{j}}{w-y_{j}}+\rme^{(L-N)\gamma}\Big(\frac{1-w}{1-qw}\Big)^{L}\prod_{j=1}^{N}\frac{qw-y_{j}}{w-y_{j}}\;
\end{equation}
is the eigenvalue of the transfer matrix $T(w)$, a fundamental object in the algebraic formulation of Bethe ansatz, see \ref{appendix algebraic Bethe ansatz}.
\end{subsection}

\begin{subsection}{Special configurations}
In order to compute the statistics of the current for a given initial condition from the expansion of (\ref{GFcurrent[M(gamma)]}) over eigenstates, the scalar product between the initial configuration and the eigenvectors is needed. Two particularly interesting classes of initial conditions are the flat configurations, where particles are equally spaced, and the step configurations, for which particles occupy consecutive sites. Exact formulas can be written for these two cases, translating previous results for the XXZ spin chain.

Let $\mathcal{F}$ be the flat configuration of a half-filled system $N=L/2$ with particles at positions $x_{j}=2j$, $j=1,\ldots,N$. Then, the XXZ result for the N\'eel state \cite{P2014.2} implies
\begin{eqnarray}
\label{psi flat[y]}
&&\fl\hspace{10mm} \langle\psi|\mathcal{F}\rangle=\frac{(-1)^{\frac{N(N-1)}{2}}\rme^{\frac{N(N+3)\gamma}{2}}}{2^{N(N+1)}}\Bigg(\prod_{j=1}^{N}\frac{(1-y_{j})^{L+2}}{y_{j}(1-qy_{j})(1-qy_{j}^{2})}\Bigg)\\
&&\fl\hspace{25mm} \times\Bigg(\prod_{j=1}^{N}\prod_{k=j+1}^{N}\frac{1}{(y_{j}-y_{k})(1-qy_{j}y_{k})}\Bigg)
\det\Bigg(\Big(\frac{1+y_{k}}{1-y_{k}}\Big)^{2j}-\Big(\frac{1+qy_{k}}{1-qy_{k}}\Big)^{2j}\Bigg)_{j,k}\;.\nonumber
\end{eqnarray}

Let $\mathcal{S}$ be the step configuration with particles at positions $x_{j}=j$, $j=1,\ldots,N$ for arbitrary density $\rho=N/L$. Another result for the XXZ spin chain gives \cite{MC2010.1}
\begin{eqnarray}
\label{psi step[y]}
&& \langle\psi|\mathcal{S}\rangle=\rme^{-\frac{2\rmi\pi pN}{L}}(-1)^{\frac{N(N+1)}{2}}\Bigg(\prod_{j=1}^{N}\frac{(1-y_{j})^{N}}{y_{j}}\Bigg)\Bigg(\prod_{j=1}^{N}\prod_{k=j+1}^{N}\frac{1}{y_{j}-y_{k}}\Bigg)\\
&&\hspace{20mm} \times\det\Bigg(\frac{1}{(1-qy_{k})^{j}}-\frac{1}{(1-y_{k})^{j}}\Bigg)_{j,k}\;.\nonumber
\end{eqnarray}
Both (\ref{psi flat[y]}) and (\ref{psi step[y]}) already assume that the $y_{j}$'s are solution of the Bethe equations.
\end{subsection}

\begin{subsection}{Sum over configurations}
Computing current fluctuations from (\ref{GFcurrent[M(gamma)]}) requires the scalar product between $\sum_{\mathcal{C}}\langle\mathcal{C}|\rme^{-\gamma S}$ and the right eigenstates. The scalar product between left eigenstates and $\sum_{\mathcal{C}}\rme^{\gamma S}|\mathcal{C}\rangle$ is also needed for the stationary initial condition $|P_{0}\rangle\propto\sum_{\mathcal{C}}|\mathcal{C}\rangle$. These scalar products can be computed from the Slavnov determinant (\ref{Slavnov}) similarly to what is done in \cite{B2009.1} for TASEP. Indeed, in the limit where all $w_{j}$'s converge to $0$, the Bethe vectors reduce to
\begin{eqnarray}
&& |\psi_{\vec{w}}\rangle\simeq\rme^{-\frac{N(N+1)\gamma}{2}}\Big(\prod_{j=1}^{N}\big(w_{j}(1-q^{j})\big)\Big)\rme^{\gamma S}\sum_{\mathcal{C}}|\mathcal{C}\rangle\\
&&\langle\psi_{\vec{w}}|\simeq\rme^{(NL-\frac{N(N-1)}{2})\gamma}\Big(\prod_{j=1}^{N}(1-q^{j})\Big)\sum_{\mathcal{C}}\langle\mathcal{C}|\rme^{-\gamma S}
\;,
\end{eqnarray}
and one finds
\begin{eqnarray}
\label{sum psi R}
&& \sum_{\mathcal{C}}\langle\mathcal{C}|\rme^{-\gamma S}|\psi\rangle=\rme^{-\frac{N(N+1)}{2}\gamma}\prod_{j=0}^{N-1}(1-\rme^{-L\gamma}q^{j})\\
\label{sum psi L}
&& \sum_{\mathcal{C}}\langle\psi|\rme^{\gamma S}|\mathcal{C}\rangle=\rme^{-\frac{N(N-1)}{2}\gamma}\Big(\prod_{j=1}^{N}y_{j}^{-1}\Big)\prod_{j=0}^{N-1}(\rme^{L\gamma}-q^{j})\;.
\end{eqnarray}
Furthermore, the mean value and stationary two-point function of the density can be computed by inserting in (\ref{GFcurrent[M(gamma)]}) operators $\eta_{i}$ counting the number of particles ($0$ or $1$) at site $i$. Formulas similar to (\ref{sum psi R}), (\ref{sum psi L}) are then needed with the operator $\openone_{\{\eta_{1}(\mathcal{C})=0\}}$ inserted. We conjecture
\begin{eqnarray}
\label{sum eta1=0 psi R}
&&\fl \sum_{\mathcal{C}}\openone_{\{\eta_{1}(\mathcal{C})=0\}}\langle\mathcal{C}|\rme^{-\gamma S}|\psi\rangle=\rme^{-\frac{N(N+1)}{2}\gamma}(\rme^{\frac{2\rmi\pi p}{L}-N\gamma}-\rme^{-L\gamma})\Big(\prod_{j=1}^{N-1}(1-\rme^{-L\gamma}q^{j})\Big)\\
\label{sum eta1=0 psi L}
&&\fl \sum_{\mathcal{C}}\openone_{\{\eta_{1}(\mathcal{C})=0\}}\langle\psi|\rme^{\gamma S}|\mathcal{C}\rangle=\rme^{-\frac{N(N-1)}{2}\gamma}\Big(\prod_{j=1}^{N}y_{j}^{-1}\Big)(\rme^{L\gamma}-\rme^{-\frac{2\rmi\pi p}{L}+N\gamma})\Big(\prod_{j=1}^{N-1}(\rme^{L\gamma}-q^{j})\Big)\,.
\end{eqnarray}
These expressions were first guessed using a computer algebra system for $N=1,2,3$ and $L$ arbitrary, by calculating explicitly the sum over all configurations and replacing each occurrence of $(1-qy_{j})^{L}$ by its expression coming from the Bethe equations (\ref{Bethe equations}). The formulas were then confirmed for all $\Omega=\C{L}{N}$ eigenstates of all systems with $1\leq N<L\leq9$, $q=0.1$ and a generic value for $\gamma$, by solving numerically the Bethe equations for the eigenstates using the method of \ref{appendix numerics Bethe roots}.
\end{subsection}

\end{section}

\begin{section}{Asymptotics of ASEP eigenstates}
\label{section asymptotics}
In this section, we consider the first eigenstates of ASEP, whose eigenvalues have a real part scaling as $L^{-3/2}$. We first recall the TASEP results, and then state precise conjectures for ASEP asymptotics guided by KPZ universality.

\begin{subsection}{First eigenstates of TASEP}
\label{section asymptotics TASEP}
Each eigenstate of TASEP, and also of ASEP by continuity, is characterized by a set of $N$ (half-)integers $k_{j}$, $j=1,\ldots,N$, defined in (\ref{g(yj)}). In order to study ASEP on the relaxation scale $t\sim L^{3/2}$, only the eigenstates of $M(\gamma)-(1-q)\rho(1-\rho)L\gamma\openone$ corresponding to eigenvalues with a real part $\sim L^{-3/2}$ are needed. For TASEP, such eigenstates can be described as quasiparticle-hole excitations over the Fermi sea $k_{j}^{0}=j-(N+1)/2$ representing the stationary state \cite{GS1992.1,P2014.1}, see also \cite{dGE2006.1} for the case with open boundaries. These eigenstates are denoted in the following by the subscript $r$. Each one is described by two finite sets of half integers $\mathbb{P},\mathbb{H}\subset\mathbb{Z}+\frac{1}{2}$, as
\begin{eqnarray}
&&\hspace{-12.5mm} \{k_{1},\ldots,k_{N}\}=\{k_{1}^{0},\ldots,k_{N}^{0}\}
\bigcup\Big\{\frac{N}{2}+a,a\in\mathbb{P}^{+}\Big\}
\bigcup\Big\{-\frac{N}{2}+a,a\in\mathbb{P}^{-}\Big\}\nonumber\\
&&\hspace{38mm} \;\Big\backslash\;\Big\{\frac{N}{2}+a,a\in\mathbb{H}^{-}\Big\}
\;\Big\backslash\;\Big\{-\frac{N}{2}+a,a\in\mathbb{H}^{+}\Big\}\;,
\end{eqnarray}
where $\mathbb{P}^{\pm},\mathbb{H}^{\pm}$ represent the positive and negative elements of $\mathbb{P}$ and $\mathbb{H}$. The elements of $\mathbb{P}$ and $\mathbb{H}$ represent respectively momenta of quasiparticle and hole excitations, with total momentum $p_{r}=\sum_{a\in\mathbb{P}}a-\sum_{a\in\mathbb{H}}a$. The cardinals of the sets verify the constraints $|\mathbb{P}^{+}|=|\mathbb{H}^{-}|$ and $|\mathbb{P}^{-}|=|\mathbb{H}^{+}|$. In particular, both $\mathbb{P}$ and $\mathbb{H}$ have the same cardinal $m_{r}=|\mathbb{P}|=|\mathbb{H}|$.

Many TASEP asymptotics of eigenstates are expressed in terms of the function
\begin{equation}
\label{chi}
\chi_{r}(v)=\chi_{0}(v)+\sum_{a\in\mathbb{P}}\frac{\omega_{a}^{3}(v)}{3}+\sum_{a\in\mathbb{H}}\frac{\omega_{a}^{3}(v)}{3}\;,
\end{equation}
with $\chi_{0}$ defined in terms of the Hurwitz $\zeta$ function as
\begin{equation}
\chi_{0}(v)=\frac{8\pi^{3/2}}{3}\Big(\sqrt{-\rmi}\,\zeta\big(-\frac{3}{2},\frac{1}{2}+\frac{\rmi v}{2\pi}\big)+\sqrt{\rmi}\,\zeta\big(-\frac{3}{2},\frac{1}{2}-\frac{\rmi v}{2\pi}\big)\Big)\;,
\end{equation}
and the elementary excitation with momentum $a$
\begin{equation}
\omega_{a}(v)=2\sqrt{\sgn(a)\rmi\pi}\sqrt{|a|+\sgn(a)\frac{\rmi v}{2\pi}}\;.
\end{equation}
The function $\chi_{0}$ is analytic for $v$ in $\mathbb{D}=\mathbb{C}\backslash\big(\rmi[\pi,\infty)\cup-\rmi[\pi,\infty)\big)$. This is also the case for $\omega_{a}$ when $a\in\mathbb{Z}+1/2$, and thus for $\chi_{r}$ for any eigenstate $r$. The function $\chi_{0}$ has an alternative expression as a polylogarithm, $\chi_{0}(v)=-(2\pi)^{-1/2}\Li_{5/2}(-\rme^{v})$ when $\Re\,v<0$ or $\Re\,v>0$, $|\Im\,v|<\pi$.

Asymptotics of eigenstates of $M(\gamma)$ for TASEP with finite rescaled fugacity
\begin{equation}
\label{s[gamma]}
s=\sqrt{\rho(1-\rho)}L^{3/2}\gamma\;
\end{equation}
also involve the quantities $\nu_{r}$ solution of
\begin{equation}
\label{nu}
\chi_{r}'(\nu_{r})=s\;,
\end{equation}
and the function
\begin{eqnarray}
&& D_{r}(\nu)=\frac{(\frac{\rmi\pi}{2})^{m_{r}^{2}}}{(2\pi)^{m_{r}}}\Big(\prod_{a,b\in\mathbb{P}\atop a>b}(a-b)\Big)\Big(\prod_{a,b\in\mathbb{H}\atop a>b}(a-b)\Big)\\
&&\hspace{15mm} \times\exp\Big(\lim_{\Lambda\to\infty}-m_{r}^{2}\log\Lambda+\int_{-\Lambda}^{\nu}\rmd v\,\frac{\chi_{r}''(v)^{2}}{2}\Big)\;.\nonumber
\end{eqnarray}
An alternative expression for $D_{r}(\nu)$ as a Cauchy determinant exists \cite{P2016.1}.
\end{subsection}

\begin{subsection}{Numerical conjectures for ASEP eigenstates}
From KPZ universality, the eigenvalues of ASEP should be the same as the ones of TASEP in the thermodynamic limit $L\to\infty$ with fixed density $\rho$ and rescaled fugacity $s$, up to a global factor $1-q$: $|E_{r}^{\text{ASEP}}-(1-q)E_{r}^{\text{TASEP}}|\ll\frac{1}{L^{3/2}}$. This was already shown for the gap $r=1$ when $\gamma=0$ \cite{K1995.1} and for the stationary state $r=0$ when $\gamma\sim L^{-3/2}$ \cite{LK1999.1}. For a general eigenstate of ASEP, obtained by continuity after increasing $q$ starting from the TASEP eigenstate with $r=(\mathbb{P},\mathbb{H})$, it leads to
\begin{equation}
\label{conjecture E}
\frac{E_{r}}{1-q}\simeq\rho(1-\rho)L\gamma-\frac{2\rmi\pi(1-2\rho)p_{r}}{L}+\sqrt{\rho(1-\rho)}\,\frac{\chi_{r}(\nu_{r})}{L^{3/2}}\;,
\end{equation}
with $\chi_{r}$ defined in (\ref{chi}), and $\nu_{r}$ the solution of (\ref{nu}) with rescaled fugacity (\ref{s[gamma]}).

From KPZ universality, one also expects that in the thermodynamic limit the Bethe eigenvectors of ASEP are the same as those of TASEP, up to a normalization factor $\lambda(q)$ independent of the eigenstate. From (\ref{sum psi R}) or (\ref{sum eta1=0 psi R}),
\begin{equation}
\lambda(q)=\prod_{j=1}^{\infty}(1-q^{j})=(q;q)_{\infty}\;.
\end{equation}
Then (\ref{sum psi L}) and (\ref{sum eta1=0 psi L}) suggest that the asymptotics of the product of the $y_{j}$'s is independent of $q$. From \cite{P2014.1}, $\prod_{j=1}^{N}y_{j}\simeq\rme^{L[\rho\log\rho+(1-\rho)\log(1-\rho)]}\,\rme^{\nu_{r}}$ and one finds
\begin{eqnarray}
\label{conjecture sum psi R}
&& \sum_{\mathcal{C}}\langle\mathcal{C}|\rme^{-\gamma S}|\psi\rangle\simeq\lambda(q)\,\rme^{-\frac{\rho^{2}s\sqrt{L}}{2\sqrt{\rho(1-\rho)}}}\frac{s}{\sqrt{\rho(1-\rho)}\sqrt{L}}\\
&& \frac{1}{\Omega}\sum_{\mathcal{C}}\langle\psi|\rme^{\gamma S}|\mathcal{C}\rangle\simeq\lambda(q)\,\rme^{\frac{\rho(2-\rho)s\sqrt{L}}{2\sqrt{\rho(1-\rho)}}}\sqrt{2\pi}\,s\,\rme^{-\nu_{r}}\\
&& \sum_{\mathcal{C}}\openone_{\{\eta_{1}(\mathcal{C})=0\}}\langle\mathcal{C}|\psi\rangle_{\gamma=0}\simeq\lambda(q)\,\frac{2\rmi\pi p_{r}}{L}\\
&& \frac{1}{\Omega}\sum_{\mathcal{C}}\openone_{\{\eta_{1}(\mathcal{C})=0\}}\langle\psi|\mathcal{C}\rangle_{\gamma=0}\simeq\lambda(q)\,\frac{2\rmi\pi p_{r}\sqrt{2\pi}\sqrt{\rho(1-\rho)}\,\rme^{-\nu_{r}}}{\sqrt{L}}\;.
\end{eqnarray}

TASEP results also lead to a conjecture for the asymptotics of the norm (\ref{norm[y]}) of Bethe eigenstates:
\begin{equation}
\label{conjecture norm}
\langle\psi_{r}|\psi_{r}\rangle\simeq
\lambda(q)^{2}\frac{\rme^{\sqrt{\rho(1-\rho)}\,s\,\sqrt{L}}}{\sqrt{\rho(1-\rho)}\sqrt{L}}\,\frac{\chi_{r}''(\nu_{r})}{D_{r}(\nu_{r})^{2}}\;.
\end{equation}
Finally, TASEP asymptotics for the components of eigenstates corresponding to a flat configuration (at half-filling $\rho=1/2$) and to a step configuration (at arbitrary filling $\rho$ with sites $1$ through $N$ occupied) lead to
\begin{eqnarray}
&& \langle\psi_{r}|\mathcal{F}\rangle\simeq\openone_{\{\mathbb{P}=\mathbb{H}\}}\lambda(q)\rme^{s\sqrt{L}/4}\frac{\rmi^{m_{r}}\rme^{-\nu_{r}/4}}{(1+\rme^{-\nu_{r}})^{1/4}D_{r}}\;
\end{eqnarray}
and
\begin{equation}
\label{conjecture step}
\langle\psi_{r}|\mathcal{S}\rangle\simeq\lambda(q)\rme^{-2\rmi\pi p_{r}\rho}\;.
\end{equation}
\end{subsection}

\begin{subsection}{Current fluctuations}
\begin{figure}
  \begin{center}
    \hspace*{-6mm}
    \begin{tabular}{lll}
      \begin{tabular}{c}
        \includegraphics[width=75mm]{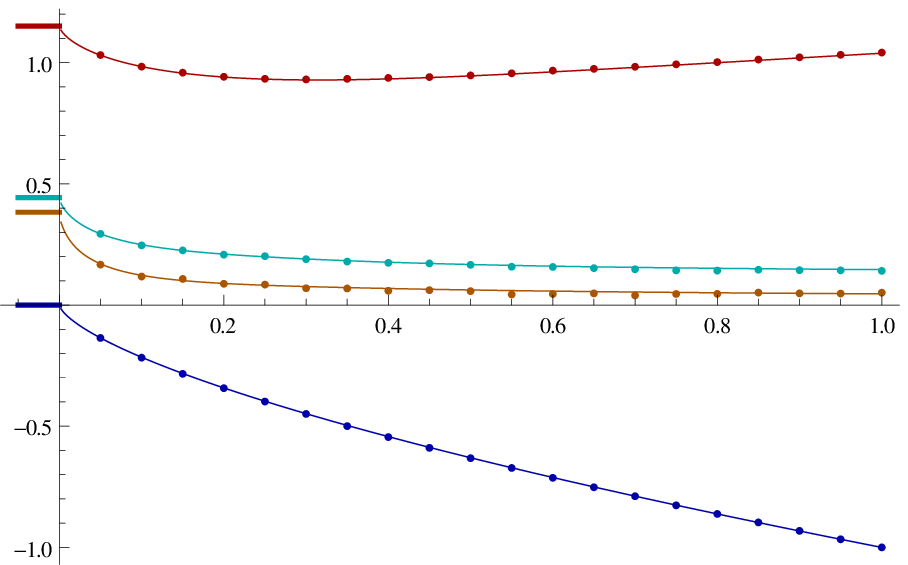}
        \begin{picture}(0,0)
          \put(-3,16.5){\small$\tau$}
          \put(-50,31){\small Stat \qquad $q=0$}
        \end{picture}
      \end{tabular}
      &\hspace*{-6mm}&
      \begin{tabular}{c}
        \includegraphics[width=75mm]{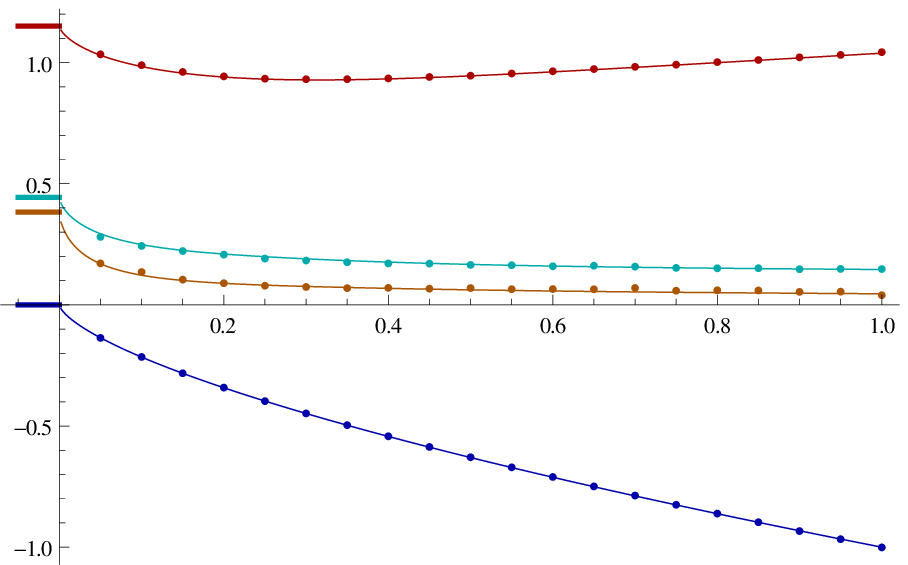}
        \begin{picture}(0,0)
          \put(-3,16.5){\small$\tau$}
          \put(-50,31){\small Stat \qquad $q=0.5$}
        \end{picture}
      \end{tabular}\\\\
      \begin{tabular}{c}
        \includegraphics[width=75mm]{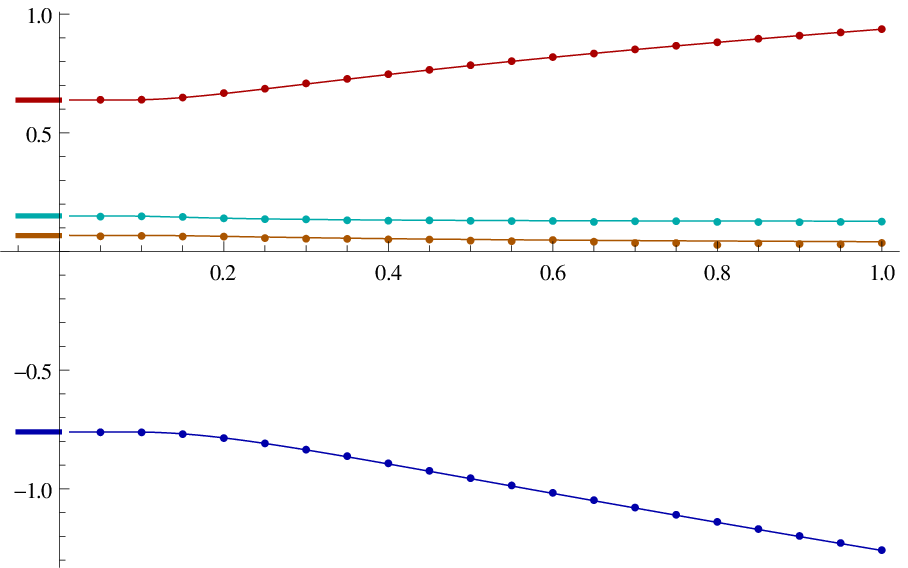}
        \begin{picture}(0,0)
          \put(-3,21.5){\small$\tau$}
          \put(-50,32){\small Flat \qquad $q=0$}
        \end{picture}
      \end{tabular}
      &\hspace*{-6mm}&
      \begin{tabular}{c}
        \includegraphics[width=75mm]{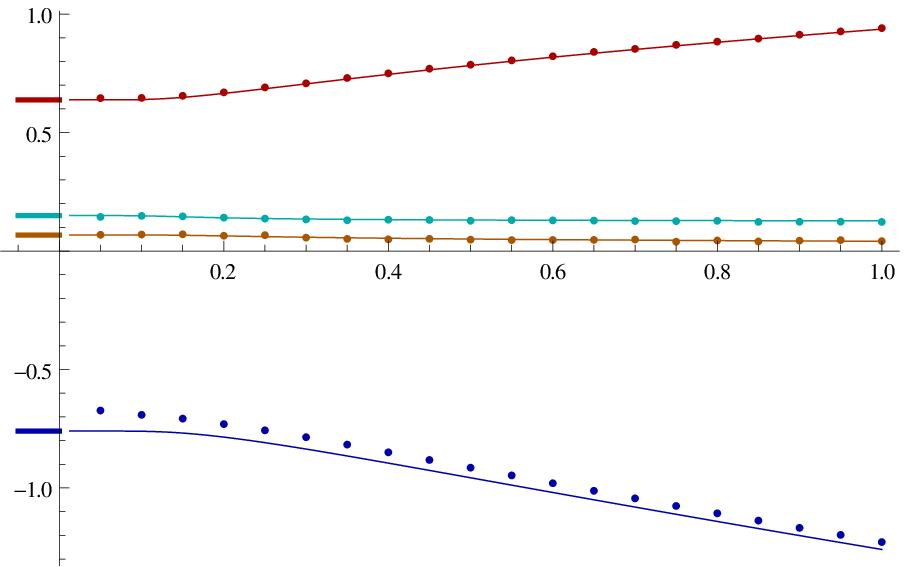}
        \begin{picture}(0,0)
          \put(-3,21.5){\small$\tau$}
          \put(-50,32){\small Flat \qquad $q=0.5$}
        \end{picture}
      \end{tabular}\\\\
      \begin{tabular}{c}
        \includegraphics[width=75mm]{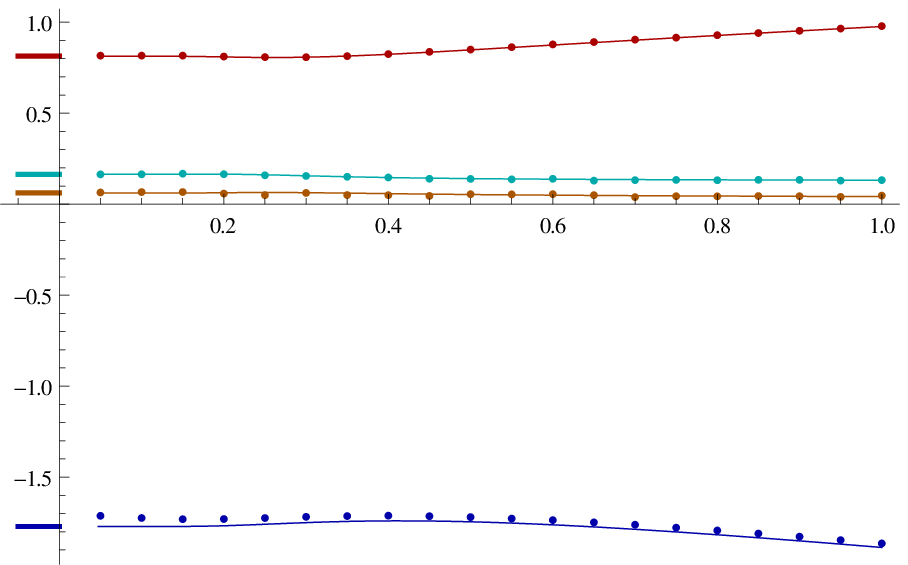}
        \begin{picture}(0,0)
          \put(-3,25){\small$\tau$}
          \put(-50,35){\small Step \qquad $q=0$}
        \end{picture}
      \end{tabular}
      &\hspace*{-6mm}&
      \begin{tabular}{c}
        \includegraphics[width=75mm]{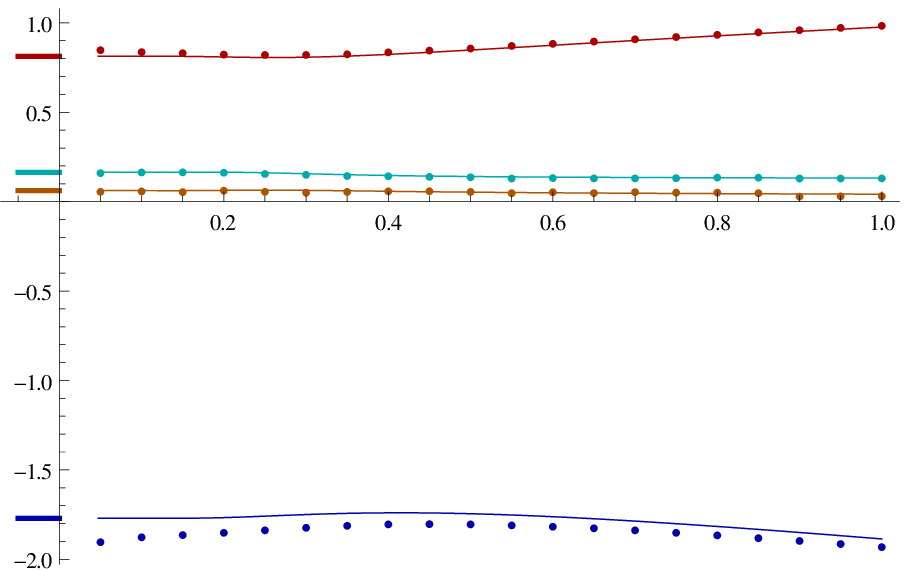}
        \begin{picture}(0,0)
          \put(-3,26){\small$\tau$}
          \put(-50,35){\small Step \qquad $q=0.5$}
        \end{picture}
      \end{tabular}
    \end{tabular}
  \end{center}
  \caption{First cumulants of the current $\langle(-\xi_{\tau}/\tau^{1/3})^{k}\rangle_{c}$ plotted as a function of the rescaled time $\tau$, for TASEP (left) and ASEP at $q=0.5$ (right), with stationary initial condition (top row), flat initial condition (middle row) and step initial condition at position $x=0$ (bottom row). In all the plots, the cumulants correspond from top to bottom to $k=2,3,4,1$. The solid lines, computed from (\ref{xi stat})-(\ref{xi step}), are independent of the asymmetry $q$. The marks on the left are the respective Tracy-Widom values for the cumulants. The dots are results of simulations with $N=1024$ particles on $L=2048$ sites averaged over $5\times10^{5}$ independent realizations.}
  \label{fig cumulants xi}
\end{figure}
The previous asymptotics lead to exact expressions for the current fluctuations in the large $L$ limit with fixed rescaled time
\begin{equation}
\tau=\frac{(1-q)t}{2L^{3/2}}\;
\end{equation}
and fixed density of particles $\rho=N/L$. In order to avoid the need of a moving reference frame, required by the typical velocity $1-2\rho$ of density fluctuations, we consider only the half-filled case $\rho=1/2$.

We define the current fluctuations as
\begin{equation}
\xi_{\tau}=\frac{Q_{i}-(1-q)t/4-\mathcal{R}L}{\sqrt{L}/2}\;.
\end{equation}
The term $\mathcal{R}$ can be understood from Burgers' hydrodynamic evolution in the time scale $t\sim L$ \cite{P2015.3}. It is equal to $\mathcal{R}=0$ for flat and stationary initial condition and $\mathcal{R}=-|x|/2$ for step initial condition with the current $Q_{i}$ counted at site $i=(x+1/2)L$, $-1/2<x<1/2$.

Expanding (\ref{GFcurrent[M(gamma)]}) over eigenstates and gathering all the asymptotics, one finds for the generating function of current fluctuations
\begin{eqnarray}
\label{xi stat}
&& \langle\rme^{s\xi_{\tau}}\rangle_{\text{stat}}=\sqrt{2\pi}s^{2}\sum_{r}\frac{D_{r}^{2}(\nu_{r})\,\rme^{\tau\chi_{r}(\nu_{r})}}{\rme^{\nu_{r}}\chi_{r}''(\nu_{r})}\\
\label{xi flat}
&& \langle\rme^{s\xi_{\tau}}\rangle_{\text{flat}}=s\sum_{r}\openone_{\{\mathbb{P}=\mathbb{H}\}}\frac{\rmi^{m_{r}}D_{r}(\nu_{r})\,\rme^{\tau\chi_{r}(\nu_{r})}}{\rme^{\nu_{r}/4}(1+\rme^{-\nu_{r}})^{1/4}\chi_{r}''(\nu_{r})}\\
&& \label{xi step}
\langle\rme^{s\xi_{\tau}}\rangle_{\text{step}}=s\sum_{r}\rme^{2\rmi\pi p_{r}x}\,\frac{D_{r}^{2}(\nu_{r})\,\rme^{\tau\chi_{r}(\nu_{r})}}{\chi_{r}''(\nu_{r})}\;.
\end{eqnarray}
The first cumulants of $\xi_{\tau}$ corresponding to (\ref{xi stat})-(\ref{xi step}) are plotted in figure \ref{fig cumulants xi} along with results from simulations of TASEP and ASEP with $q=1/2$. The agreement is excellent for TASEP, except for the mean value in the step case due to $1/\sqrt{L}$ finite size corrections instead of $1/L$ for all the other cumulants. The agreement is also good for ASEP, except for the mean value with flat and step initial condition.

The functions $\chi_{r}$, $D_{r}$ and the quantities $\nu_{r}$ are defined in section \ref{section asymptotics TASEP}. At long time (\ref{xi stat})-(\ref{xi step}) imply that the large deviation function of $\xi_{\tau}$ is the Derrida-Lebowitz function, while at short time numerical evaluations of (\ref{xi stat})-(\ref{xi step}) indicate that the statistics of $\tau^{-1/3}\xi_{\tau}$ is described by a Tracy-Widom distribution, in agreement with the results on the infinite line \cite{P2016.1}.

\begin{figure}
  \begin{center}
    \hspace*{-6mm}
    \begin{tabular}{lll}
      \begin{tabular}{c}
        \includegraphics[width=75mm]{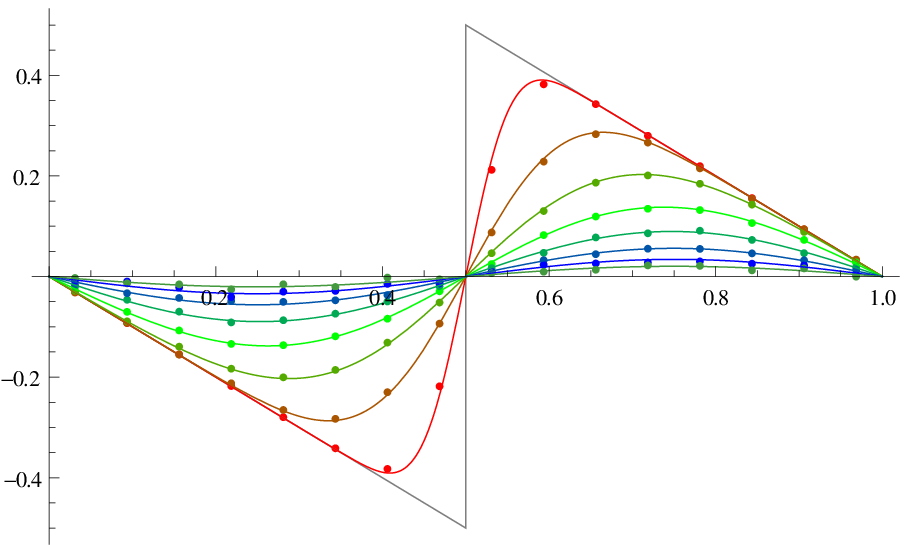}
        \begin{picture}(0,0)
          \put(-3,24){\small$x$}
          \put(-60,35){\small $q=0$}
        \end{picture}
      \end{tabular}
      &\hspace*{-6mm}&
      \begin{tabular}{c}
        \includegraphics[width=75mm]{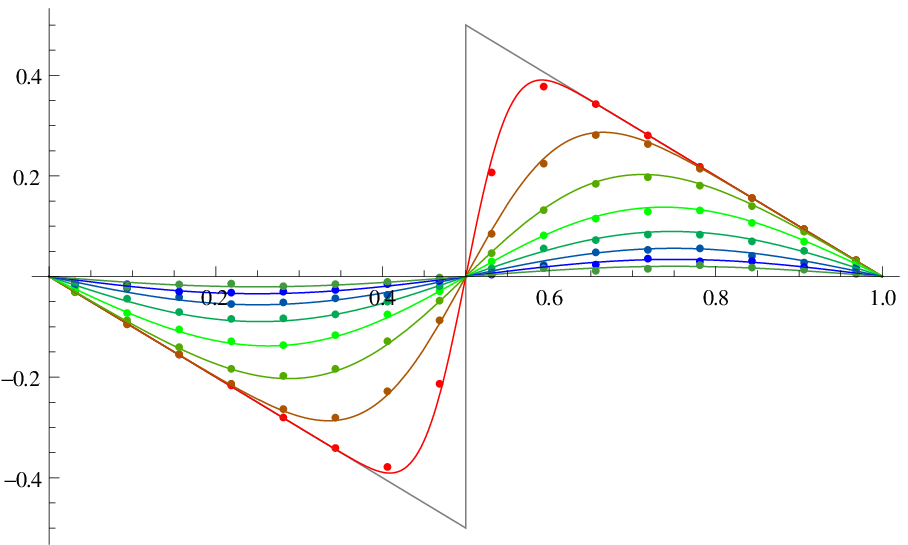}
        \begin{picture}(0,0)
          \put(-3,24){\small$x$}
          \put(-60,35){\small $q=0.5$}
        \end{picture}
      \end{tabular}
    \end{tabular}
  \end{center}
  \caption{Average density of particles $2\tau\langle\sigma(x,\tau)\rangle$ with step initial condition, plotted for TASEP (left) and ASEP at $q=0.5$ (right) as a function of the position $x$, for times $\tau=0.05,0.1,\ldots,0.4$ from largest to smallest amplitude. The shock obtained at short time is represented in grey. The solid lines, computed from (\ref{sigma step}), are independent of the asymmetry $q$. The dots are results of simulations with $N=1024$ particles on $L=2048$ sites averaged over $5\times10^{5}$ independent realizations and $128$ consecutive sites.}
  \label{fig sigma1}
\end{figure}
The asymptotics of eigenstates also give the mean value and stationary two-point function of the density by setting $\gamma=0$ in (\ref{GFcurrent[M(gamma)]}). By translation invariance, $\langle\eta_{i}\rangle_{\text{stat}}=\langle\eta_{i}\rangle_{\text{flat}}=1/2$. For step initial condition with sites $1$ through $N$ occupied, the density fluctuations $\sigma(x,\tau)=2\sqrt{L}(\eta_{i}-\tfrac{1}{2})$ at site $i=(\rho+x)L$ are equal on average to
\begin{equation}
\label{sigma step}
\langle\sigma(x,\tau)\rangle_{\text{step}}=-2\rmi\pi\sum_{r>0}\rme^{2\rmi\pi p_{r}x}\,\frac{p_{r}D_{r}^{2}(\nu_{r})\,\rme^{\tau\chi_{r}(\nu_{r})}}{\chi_{r}''(\nu_{r})}\;.
\end{equation}
The sum is over all eigenstates $r=(\mathbb{P},\mathbb{H})$ except the stationary eigenstate. The exact expression (\ref{sigma step}), plotted in figure \ref{fig sigma1}, agrees very well with simulations of TASEP and ASEP.

\begin{figure}
  \begin{center}
    \hspace*{-6mm}
    \begin{tabular}{lll}
      \begin{tabular}{c}
        \includegraphics[width=75mm]{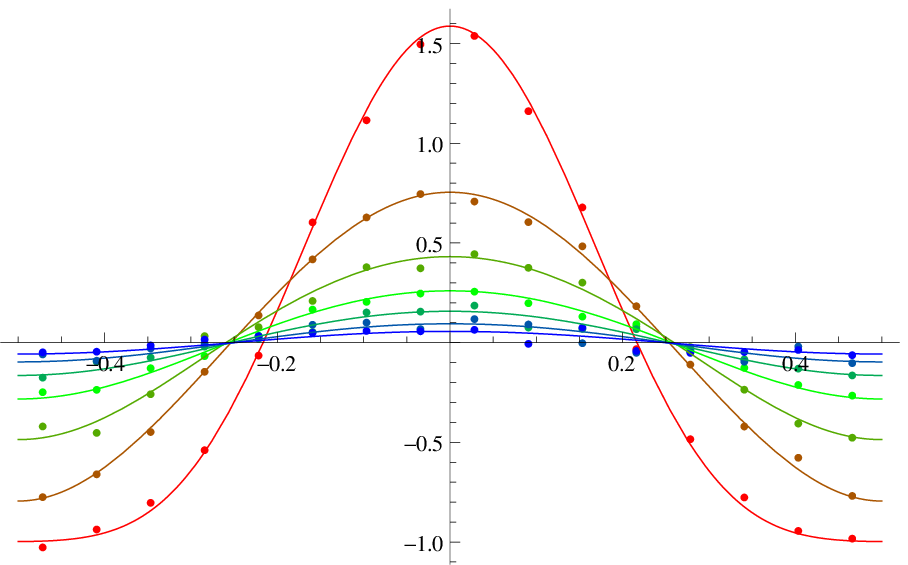}
        \begin{picture}(0,0)
          \put(-3,20){\small$x$}
          \put(-70,35){\small $q=0$}
        \end{picture}
      \end{tabular}
      &\hspace*{-6mm}&
      \begin{tabular}{c}
        \includegraphics[width=75mm]{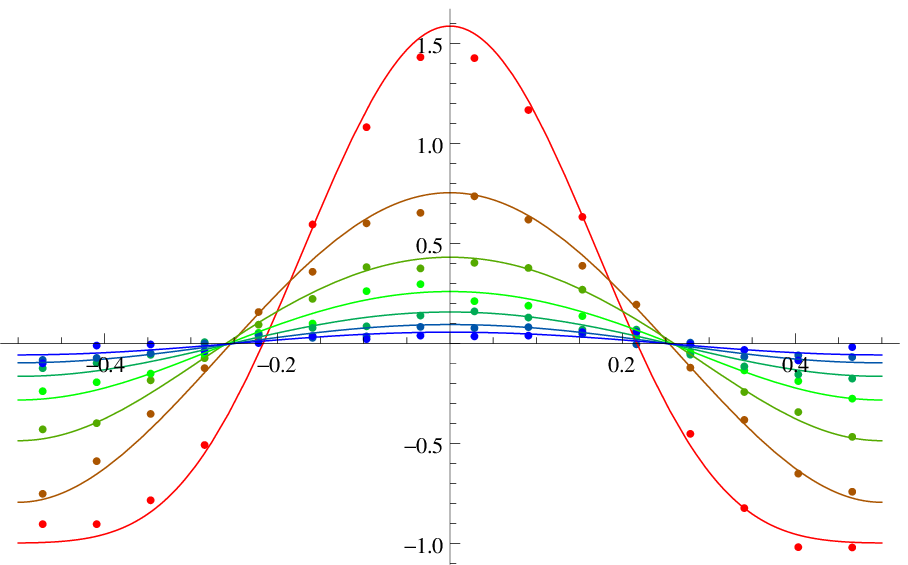}
        \begin{picture}(0,0)
          \put(-3,20){\small$x$}
          \put(-70,35){\small $q=0.5$}
        \end{picture}
      \end{tabular}
    \end{tabular}
  \end{center}
  \caption{Stationary two-point function of the density of particles $\langle\sigma(0,0)\sigma(x,\tau)\rangle$, plotted for TASEP (left) and ASEP at $q=0.5$ (right) as a function of the position $x$, for times $\tau=0.04,0.08,\ldots,0.28$ from largest to smallest amplitude. The solid lines, computed from (\ref{sigma2 stat}), are independent of the asymmetry $q$. The dots are results of simulations with $N=256$ particles on $L=512$ sites averaged over $10^{7}$ independent realizations and $32$ consecutive sites.}
  \label{fig sigma2}
\end{figure}
Similarly, one finds for the stationary two-point function of the density
\begin{equation}
\label{sigma2 stat}
\langle\sigma(0,0)\sigma(x,\tau)\rangle_{\text{stat}}=-(2\pi)^{5/2}\sum_{r>0}\rme^{2\rmi\pi p_{r}x}\,\frac{p_{r}^{2}D_{r}^{2}(\nu_{r})\,\rme^{\tau\chi_{r}(\nu_{r})}}{\rme^{\nu_{r}}\chi_{r}''(\nu_{r})}\;.
\end{equation}
The stationary two-point function is plotted in figure \ref{fig sigma2}. The agreement with simulations is good, although noticeably worse than for the one-point function due to the smaller system size studied, which was imposed by the much larger number of realizations needed in order to compute accurately the average.

The expressions (\ref{xi stat})-(\ref{sigma2 stat}) are identical to the ones obtained for TASEP in \cite{P2016.1}. They look quite similar to discretized functional integrals with the action $\int_{-\infty}^{\nu}\rmd v(\varphi'(v)^{2}+\tau\varphi(v))$ of a scalar field in a linear potential. The kinetic part of the action is hidden in $D_{r}^{2}(\nu)$. The discrete realizations $\varphi_{r}$ of the field $\varphi$ are related to the functions $\chi_{r}$ by $\varphi_{r}(v)=\chi_{r}'(v)$. Since the value of $\varphi_{r}$ at the upper limit of the integral in the action is equal to $\varphi_{r}(\nu_{r})=s$, one can identify $\varphi_{r}$ as a field conjugate to the current.
\end{subsection}

\end{section}

\begin{section}{Extrapolation methods for asymptotics of eigenstates}
\label{section numerics}
Exact calculations for TASEP \cite{P2014.1,P2015.2} indicate that quantities such as eigenvalues, normalization of Bethe states, and components of Bethe states for simple configurations have a clean asymptotic expansion with exponent $\theta=1/2$
\begin{equation}
\label{expansion powers z^theta}
\sum_{k=0}^{\infty}a_{k}z^{\theta k}\;
\end{equation}
in the variable $z=1/L$. Assuming that the same is true for ASEP, which is supported by the exact calculation \cite{K1995.1} of the gap, powerful extrapolation methods can be used to extract numerical evaluations with high accuracy of large $L$ asymptotics from the knowledge of a few finite size values. We believe that extrapolation methods should also be very successful in many other integrable models, due to the the availability of finite size expressions that can be evaluated numerically with high precision for moderately large systems combined with the existence of clean asymptotic expansions (see \textit{e.g.} \cite{K2015.1} for an example in the XXZ spin chain, where expansions with $\theta=1$ are found).

\begin{subsection}{Richardson extrapolation}
Expansions of the form (\ref{expansion powers z^theta}) allow an extremely precise evaluation of the leading term $a_{0}$ from the knowledge of a few values $f(z_{1})$, \ldots, $f(z_{n})$ with $0<z_{1}<\ldots<z_{n}$, by removing step by step the contributions of $a_{1}$, $a_{2}$, \ldots, $a_{n-1}$. Starting from the values $f_{m}^{1}=f(z_{m})$, $m=1,\ldots,n$, one builds the triangle
\begin{equation}
\begin{array}{lllll}
  f_{1}^{1} & f_{2}^{1} & \ldots & f_{n-1}^{1} & f_{n}^{1} \\
  f_{1}^{2} & f_{2}^{2} & \ldots & f_{n-1}^{2} & \\
  \ldots & \ldots & \ldots & & \\
  f_{1}^{n-1} & f_{2}^{n-1} & & & \\
  f_{1}^{n} & & & &
\end{array}
\end{equation}
by some iteration procedure, defining $f_{m}^{p}$ in terms of some $f_{j}^{k}$ with $(j,k)$ close to $(m,p)$ and $k<p$. The extrapolated value of $a_{0}$ is then $f_{1}^{n}$, and a possible estimation for the order of magnitude of the error is $|f_{1}^{n}-f_{1}^{n-1}|+|f_{1}^{n}-f_{2}^{n-1}|+|f_{1}^{n-1}-f_{2}^{n-1}|$.

This procedure is called Richardson extrapolation. Several versions exist depending on the precise choice of the iteration. The Aitken-Neville step
\begin{equation}
f_{m}^{p}=f_{m+1}^{p-1}+\frac{f_{m+1}^{p-1}-f_{m}^{p-1}}{\big(\frac{z_{m+1}}{z_{m+p+1}}\big)^{\theta}-1}
\end{equation}
corresponds to polynomial extrapolation: it leads to $f_{1}^{n}=P(0)$ with $P$ the unique interpolating polynomial of degree $n-1$ such that $P(z_{j})=f(z_{j})$, $j=1,\ldots,n$. The Bulirsch-Stoer method \cite{HS1988.1}, which we use in this paper, has
\begin{equation}
f_{m}^{p}=f_{m+1}^{p-1}+\frac{f_{m+1}^{p-1}-f_{m}^{p-1}}{\big(\frac{z_{m+1}}{z_{m+p+1}}\big)^{\theta}\Big(1-\frac{f_{m+1}^{p-1}-f_{m}^{p-1}}{f_{m+1}^{p-1}-f_{m+1}^{p-2}}\Big)-1}
\end{equation}
with $f_{m}^{0}=0$. The iteration corresponds to rational extrapolation: it leads to $f_{1}^{n}=R(0)$ where $R$ is the unique interpolating rational function with numerator and denominator of respective degrees $\lfloor(n-1)/2\rfloor$ and $\lfloor n/2\rfloor$ such that $R(z_{j})=f(z_{j})$, $j=1,\ldots,n$. Rational extrapolation gives usually better results, especially when $f$ has singularities close to $0$.

Both iteration procedures require for large $n$ the knowledge of the $f(z_{j})$ with high precision. Typically, one needs better precision than the standard $16$ digits double precision in order for the result of the extrapolation to improve when increasing $n$ to values larger than $10$ or $15$.
\end{subsection}

\begin{subsection}{Application to asymptotics of ASEP eigenstates}
\begin{table}
  \begin{center}
    \begin{tabular}{lll}
      $L$ & $\hspace{10mm}$finite size & $\hspace{10mm}$extrapolation\\
      $10$ & $1.26682 + 0.65827 \rmi$ & $1. + 0. \rmi$\\
      $20$ & $1.26803 + 0.64323 \rmi$ & $1.3 + 0.6 \rmi$\\
      $30$ & $1.26771 + 0.63903 \rmi$ & $1.3 + 0.6 \rmi$\\
      $40$ & $1.26709 + 0.63737 \rmi$ & $1.2602 + 0.6370 \rmi$\\
      $50$ & $1.26648 + 0.63661 \rmi$ & $1.26017 + 0.63698 \rmi$\\
      $60$ & $1.26593 + 0.63622 \rmi$ & $1.260174 + 0.636979 \rmi$\\
      $70$ & $1.26546 + 0.63601 \rmi$ & $1.2601739 + 0.6369793 \rmi$\\
      $80$ & $1.26506 + 0.63590 \rmi$ & $1.26017394 + 0.63697930 \rmi$\\
      $90$ & $1.26471 + 0.63585 \rmi$ & $1.2601739429 + 0.6369792988 \rmi$\\
      $100$ & $1.26440 + 0.63582 \rmi$ & $1.26017394293 + 0.63697929878 \rmi$\\
      $110$ & $1.26414 + 0.63582 \rmi$ & $1.260173942929 + 0.636979298781 \rmi$\\
      $120$ & $1.26390 + 0.63583 \rmi$ & $1.260173942929 + 0.636979298781 \rmi$\\
      $130$ & $1.26370 + 0.63585 \rmi$ & $1.260173942929154 + 0.636979298780668 \rmi$\\
      $140$ & $1.26351 + 0.63587 \rmi$ & $1.2601739429291542 + 0.6369792987806681 \rmi$\\
      $150$ & $1.26334 + 0.63589 \rmi$ & $1.26017394292915417 + 0.63697929878066807 \rmi$\\
      $160$ & $1.26319 + 0.63591 \rmi$ & $1.2601739429291541672 + 0.6369792987806680721 \rmi$\\
      $170$ & $1.26306 + 0.63594 \rmi$ & $1.26017394292915416715 + 0.63697929878066807205 \rmi$\\
      $180$ & $1.26293 + 0.63596 \rmi$ & $1.260173942929154167146 + 0.636979298780668072053 \rmi$\\
      $190$ & $1.26282 + 0.63599 \rmi$ & $1.2601739429291541671458 + 0.6369792987806680720532 \rmi$\\
    \end{tabular}
  \end{center}
  \caption{Extrapolation for the scalar product of the stationary eigenstate $\sqrt{L}\,\rme^{-s\sqrt{L}/2}\langle\psi_{0}|\psi_{0}\rangle$ at density $\rho=1/2$, asymmetry $q=0.1$ and rescaled fugacity $s=0.2+\rmi$. The second column corresponds to finite size values. The element at a given row $L$ of the last column is the result of rational extrapolation applied to all finite size values with even system sizes $\leq L$ computed with $1000$ digits precision. A numerical evaluation of (\ref{conjecture norm}) gives $\sqrt{L}\,\rme^{-s\sqrt{L}/2}\langle\psi_{0}|\psi_{0}\rangle\approx 1.26017394292915416714584555428+0.6369792987806680720531597991129\rmi$.}
  \label{table extrapolation}
\end{table}
We used the rational extrapolation method in order to check the conjectures of section \ref{section asymptotics} for the asymptotics of ASEP eigenstates. All the computations were done for the generic value of the fugacity $s=0.2+\rmi$.

Numerics for finite systems with $\approx1000$ digits precision were obtained from the resolution of the Bethe equations using the procedure detailed in \ref{appendix numerics Bethe roots}. We used the variant with Newton's method, which seemed generally faster for our computations, especially for larger systems. This might however be due to our particular implementation, and more advanced methods to solve the system of differential equations in the other variant discussed in \ref{appendix numerics Bethe roots} might be faster, in conjunction with Newton's method in the end to get arbitrary high accuracy.

The finite size calculations were performed at density $\rho=1/2$, $\rho=1/3$ and $\rho=1/4$ with asymmetry $q=0.1$ (respectively $q=0.5$), for all $18$ (resp. $8$) eigenstates $r=(\mathbb{P},\mathbb{H})$ with $M=\sum_{a\in\mathbb{P}}|a|+\sum_{a\in\mathbb{H}}|a|\leq3$ (resp. $M\leq2$). The system sizes considered were $L=\rho^{-1}(M+1),\ldots,L_{\text{max}}-\rho^{-1},L_{\text{max}}$ with $L_{\text{max}}$ equal for density $\rho=1/2,\,1/3,\,1/4$ to $L_{\text{max}}=300,390,520$ (resp. $L_{\text{max}}=260,390,520$).

The asymptotics (\ref{conjecture E}) for the eigenvalues was checked by subtracting from the finite size values the first two terms of the asymptotics and multiplying everything by $L^{3/2}$ before using the extrapolation method. The asymptotics (\ref{conjecture sum psi R})-(\ref{conjecture step}) were checked by computing the ratio between the finite size values and the divergent factors of the asymptotics before using the extrapolation method. The finite size values are computed for all system sizes up to the maximal system size considered. The result of the extrapolation method truncated at the estimation of the error is then compared with the conjectured asymptotics. A perfect match was found within the estimated error of the extrapolation method, which corresponds to at least $22$ digits precision in all cases with $q=0.1$ and to at least $8$ digits precision in all cases with $q=0.5$.
\end{subsection}

\begin{subsection}{Application to the spectral gap of WASEP: double extrapolation}
The extrapolation method can also be useful in cases where asymptotics are not known. We illustrate this here in the weakly asymmetric regime (WASEP), where
\begin{equation}
1-q=\frac{\mu}{\sqrt{L}}\;
\end{equation}
with fixed $\mu$. This regime corresponds to the crossover between KPZ and equilibrium fluctuations, for which few analytical results have been obtained so far with periodic boundary conditions. We focus on half-filled case $\rho=1/2$ with fugacity $\gamma=0$ and consider the spectral gap $E_{1}$, equal to the non-zero eigenvalue with largest real part. One expects
\begin{equation}
\label{gap WASEP rescaled}
E_{1}\simeq\frac{e_{1}(\mu)}{L^{2}}\;.
\end{equation}
The function $e_{1}$ is plotted in figure (\ref{fig e WASEP}). For the symmetric exclusion process $q=1$, $E_{1}\simeq-4\pi^{2}/L^{2}$ \cite{GS1992.1}, thus $e_{1}(0)=-4\pi^{2}$. Extrapolation of finite size numerics with $L=2,4,\ldots,70$ for $\mu=1,0.1,0.01,0.001$ strongly indicates $e_{1}(\mu)\simeq-4\pi^{2}-\mu^{2}/2$ near $\mu=0$. This suggests an expansion of the form
\begin{equation}
e_{1}(\mu)=\sum_{j=0}^{\infty}d_{j}\,\mu^{2j}\;,
\end{equation}
with $d_{0}=-4\pi^{2}$ and $d_{1}=-1/2$. A precise numerical value of $d_{2}$ can be obtained using the extrapolation method twice, by first extrapolating finite size values of the gap in the variable $L$ with exponent $\theta=1/2$ for several values of $\mu$, and then extrapolating $(e_{1}(\mu)-d_{0}-d_{1}\mu^{2})/\mu^{4}$ in the variable $\mu$ with exponent $\theta=2$. We find $d_{2}\approx0.0040840926890$. Using an integer relation algorithm to seek a linear combination with small integer coefficients of $d_{2}$, $1$ and $\pi^{-2}$ equal to zero, we recognize that $d_{2}$ is equal to $1/96-1/(16\pi^{2})$. One can then iterate and guess exact expressions for the first $d_{j}$'s. Computing the finite size value of the gap for $L=2,4,\ldots,140$ and $\mu=0.01,0.02,\ldots,1.39,1.4$, we obtain
\begin{eqnarray}
\label{dj gap WASEP}
&& d_{0}=-4\pi^{2}\\
&& d_{1}=-\frac{1}{2}\nonumber\\
&& d_{2}=\frac{1}{96}-\frac{1}{16\pi^2}\nonumber\\
&& d_{3}=-\frac{1}{11520}+\frac{1}{384\pi^2}-\frac{7}{256\pi^4}\\
&& d_{4}=\frac{11}{3870720}+\frac{1}{30720\pi^2}+\frac{7}{4096\pi^4}-\frac{77}{4096\pi^6}\nonumber\\
&& d_{5}=-\frac{23}{185794560}-\frac{11}{7741440\pi^2}+\frac{3}{163840\pi^4}+\frac{77}{49152\pi^6}-\frac{1093}{65536\pi^8}\nonumber\\[5pt]
&& d_{6}\approx1.2205309593117315487139207309742230\times10^{-8}\;.\nonumber
\end{eqnarray}
The first $d_{j}$'s, $j\geq1$ have some structure (the denominators are highly factorizable, the second term of $d_{j+1}$ is equal to $\pm j/(8\pi^{2})$ times the first term of $d_{j}$, the next-to-last term of $d_{j+1}$ is equal to $-j/48$ times last term of $d_{j+1}$). We were however not able to find a full recursion formula to build the $d_{j}$'s.

At large $\mu$, the ASEP result $E_{1}=(1-q)e_{1}^{\text{TASEP}}/L^{3/2}$ leads to $e_{1}(\mu)\simeq e_{1}^{\text{TASEP}}\mu$ with $e_{1}^{\text{TASEP}}\approx-6.50918933798847$. Higher order corrections $e_{1}(\mu)\simeq e_{1}^{\text{TASEP}}\mu+c_{1}\mu^{-2}+c_{2}\mu^{-4}$ with $c_{1}=-128\pi^{2}\zeta(3)$ and $c_{2}\approx119833.6381$ were obtained in \cite{K1995.1}. The double extrapolation for $\mu=1.5,1,\ldots,14.5,15$ and even system sizes $\mu^{2}<L\leq L_{\text{max}}$ with $L_{\text{max}}$ increasing between $L_{\text{max}}=74$ at $\mu=1.5$ and $L_{\text{max}}=280$ at $\mu=15$ gives $\lim_{\mu\to\infty}\mu^{4}\big[e_{1}(\mu)-(e_{1}^{\text{TASEP}}\mu+c_{1}\mu^{-2})\big]\approx1.198\times10^{5}$, in reasonable agreement with $c_{2}$.
\begin{figure}
  \begin{center}
    \begin{tabular}{c}
      \includegraphics[width=100mm]{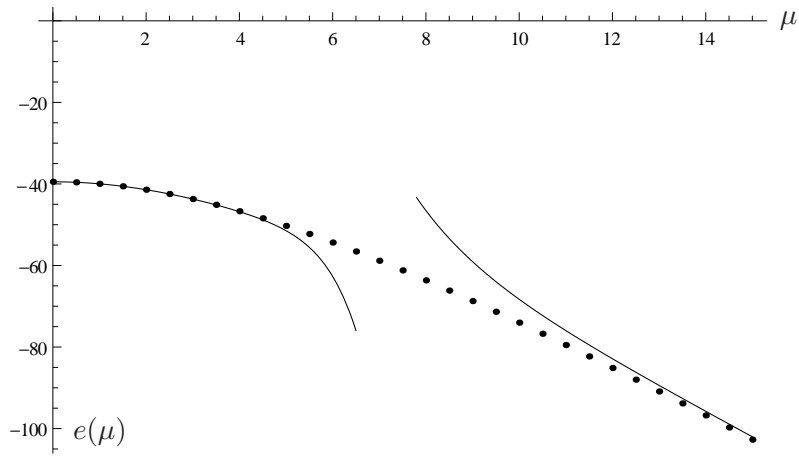}
      \begin{picture}(0,0)
        \put(0,57){\small$\mu$}
        \put(-93,2){\small$e(\mu)$}
      \end{picture}
    \end{tabular}
  \end{center}
  \caption{Spectral gap for ASEP with weak asymmetry $1-q=\mu/\sqrt{L}$ (WASEP) in the limit $L\to\infty$. The dots are the result of the extrapolation method applied on finite size Bethe ansatz numerics. The solid curve at small $\mu$ is $\sum_{j=0}^{5}d_{j}\,\mu^{2j}$ with the $d_{j}$'s given by (\ref{dj gap WASEP}). The solid curve at large $\mu$ is $e_{1}^{\text{TASEP}}\mu+c_{1}\mu^{-2}+c_{2}\mu^{-4}$.}
  \label{fig e WASEP}
\end{figure}
\end{subsection}

\end{section}

\begin{section}{Conclusions}
We have shown in this paper that ASEP and TASEP have exactly the same fluctuations on the relaxation scale, which confirms that KPZ universality extends to the whole crossover between short time transient fluctuations and stationary fluctuations. This result follows from high precision evaluations of asymptotics of eigenvalues and eigenvectors using extrapolation methods on finite size Bethe ansatz numerics. The eigenvalues and eigenvectors of ASEP and TASEP are conjectured to be essentially the same, which leads to several formulas (\ref{conjecture E}), (\ref{conjecture sum psi R})-(\ref{conjecture step}) for asymptotics of ASEP eigenstates.

We believe that proving these conjectured asymptotics would be a necessary first step in probing the crossover between equilibrium and KPZ fluctuations, which occurs in the weakly asymmetric regime $1-q\sim1/\sqrt{L}$. Exact calculations for ASEP are however much more complicated than their counterparts for TASEP because of more complicated Bethe ansatz equations, and because of the presence of determinants that are not of Vandermonde type in the scalar products of Bethe vectors.

The extrapolation methods used in this paper to check the asymptotics should be useful for many other integrable models, at least in order to check asymptotic expressions, or even to guess new formulas as illustrated in the case of the spectral gap for ASEP with weak asymmetry.
\end{section}

\appendix
\begin{section}{Algebraic Bethe ansatz for ASEP}
\label{appendix algebraic Bethe ansatz}
In this appendix, we summarize briefly the algebraic Bethe ansatz construction of the eigenstates of the generator of ASEP as products of creation operators acting on a reference state. For more details, we refer to \cite{GM2006.1}.

The creation operators, $B$ and $C$, are built from the monodromy matrix
\begin{equation}
\label{T monodromy}
\mathbb{T}_{a}(\mu)=\mathbb{L}_{aL}(\mu)\ldots\mathbb{L}_{a2}(\mu)\mathbb{L}_{a1}(\mu)
=\left(
  \begin{array}{cc}
    A(\mu) & B(\mu) \\
    C(\mu) & D(\mu)
  \end{array}
\right)\;.
\end{equation}
The monodromy matrix acts on the space $V_{a}\otimes V_{1}\otimes\ldots\otimes V_{L}$ with $V_{i}$, $i=1,\ldots,L$ the two-dimensional vector space corresponding to the site $i$ of an exclusion process, and $V_{a}$ an auxiliary space, also taken two-dimensional here. The space $V_{i}$ is generated by the vectors $|1\rangle_{i},|0\rangle_{i}$, corresponding respectively to an occupied site and to an empty site. The Lax operator $\mathbb{L}_{ai}(\mu)$ acts non-trivially only on the auxiliary site $a$ and on site $i$. Its local matrix in the basis $(|1\rangle_{a}\otimes|1\rangle_{i},|1\rangle_{a}\otimes|0\rangle_{i},|0\rangle_{a}\otimes|1\rangle_{i},|0\rangle_{a}\otimes|0\rangle_{i})$ is
\begin{equation}
\mathbb{L}_{ai}(\mu)=
\left(
  \begin{array}{cccc}
    1 & 0 & 0 & 0\\
    0 & \rme^{\gamma}\frac{1-\mu}{1-q\mu} & \frac{1-q}{1-q\mu} & 0\\
    0 & \frac{(1-q)\mu}{1-q\mu} & q\,\rme^{-\gamma}\frac{1-\mu}{1-q\mu} & 0\\
    0 & 0 & 0 & 1
  \end{array}
\right)\;.
\end{equation}
It can be shown that
\begin{equation}
\label{psiR algebraic}
|\psi\rangle=C(y_{1})\ldots C(y_{N})|\emptyset\rangle\;
\end{equation}
is an eigenvector of $M(\gamma)$ if the Bethe roots $y_{1}$, \ldots, $y_{N}$ are solution of the Bethe equations (\ref{Bethe equations}). The reference state $|\emptyset\rangle$ corresponds to the configuration with no particles. The left eigenvectors are built similarly, using the creation operators $B(y)$ instead,
\begin{equation}
\label{psiL algebraic}
\langle\psi|=\langle\emptyset|B(y_{1})\ldots B(y_{N})\;.
\end{equation}

A key point for the integrability of ASEP is that the Lax operators satisfy the Yang-Baxter equation
\begin{equation}
\mathbb{R}_{ab}(\mu/\nu)\mathbb{L}_{a}(\mu)\mathbb{L}_{b}(\nu)=\mathbb{L}_{a}(\nu)\mathbb{L}_{b}(\mu)\mathbb{R}_{ab}(\mu/\nu)\;,
\end{equation}
with $\mathbb{R}_{ab}(\mu)=\mathbb{P}_{ab}\mathbb{L}_{ab}(\mu)$ and $\mathbb{P}_{ab}$ the operator permuting the two auxiliary sites $a$ and $b$. By construction, the Yang-Baxter equation also holds when replacing Lax operators $\mathbb{L}$ by monodromy matrices $\mathbb{T}$, which implies that the transfer matrix
\begin{equation}
\label{T transfer}
T(\mu)=\tr_{a}\mathbb{T}_{a}(\mu)=A(\mu)+D(\mu)\;
\end{equation}
verifies the commutation relation $T(\mu)T(\nu)=T(\nu)T(\mu)$ for all $\mu$, $\nu$. The existence of this commuting family of operators, to which belongs the deformed Markov matrix $M(\gamma)=T'(0)T^{-1}(0)$, is at the heart of the integrability of ASEP. In particular, (\ref{psiR algebraic}) and (\ref{psiL algebraic}) are eigenvectors of $T(\mu)$ for all $\mu$, with eigenvalue $E(\mu)$ defined in (\ref{eigenvalue T}).

The Yang-Baxter equation gives a quadratic algebra for the operators $A$, $B$, $C$, $D$ defined in (\ref{T monodromy}), in particular $B(\mu)B(\nu)=B(\nu)B(\mu)$ and $C(\mu)C(\nu)=C(\nu)C(\mu)$, which means that the order of the operators in (\ref{psiR algebraic}) and (\ref{psiL algebraic}) does not matter: the eigenvectors are symmetric functions of the Bethe roots. Explicit expressions for the components of the eigenvectors for configurations with particles at positions $x_{j}$, $j=1,\ldots,N$, are given by the coordinate form of the Bethe ansatz (\ref{psiR coordinate}), (\ref{psiL coordinate}), see \textit{e.g.} \cite{GM2006.2} for a derivation from (\ref{psiR algebraic}), (\ref{psiL algebraic}).
\end{section}

\begin{section}{Numerical solution of the Bethe equations}
\label{appendix numerics Bethe roots}
In this appendix, we explain the method used in this paper to solve numerically the Bethe equations (\ref{Bethe equations}). The method has two steps: we first solve the Bethe equations for TASEP using their simple structure when $q=0$. We then use the TASEP result as a starting point for solving the Bethe equations of ASEP for a small value of $q$, and then iterate up to a target value with $0<q<1$. Two iteration procedures are considered, either formulating the problem as a system of differential equations obtained by taking the derivative of the Bethe equations with respect to $q$, or using multidimensional Newton's method for solving directly the Bethe equations.

\begin{subsection}{First step: TASEP}
At $q=0$, taking both sides of the Bethe equations (\ref{Bethe equations}) to the power $1/L$ gives
\begin{equation}
\label{g(yj)}
g(y_{j})=\rme^{\frac{2\rmi\pi k_{j}}{L}-b}\;,
\end{equation}
where the $k_{j}$'s, distinct modulo $L$, characterize the eigenstates. The $k_{j}$'s are integers (half-integers) if $N$ is odd (even). The function $g$ is defined by
\begin{equation}
g(y)=\frac{1-y}{y^{\rho}}\;,
\end{equation}
and $b$ is solution of
\begin{equation}
\label{b[y]}
b-\gamma=\frac{1}{L}\sum_{j=1}^{N}\log y_{j}\;.
\end{equation}
This last equation can be solved numerically for $b$ using Newton's method, repeatedly replacing $b$ by
\begin{equation}
b_{\text{new}}=b-\frac{b-\gamma-\frac{1}{L}\sum_{j=1}^{N}\log y_{j}(b)}{\frac{1}{N}\sum_{j=1}^{N}\frac{y_{j}(b)}{\rho+(1-\rho)y_{j}(b)}}\;
\end{equation}
until the desired precision is reached. At each step, the $y_{j}(b)$, $j=1,\ldots,N$ are computed by inverting $g$ in (\ref{g(yj)}) also using Newton's method: $y=g^{-1}(z)$ is obtained by repeatedly replacing $y$ with
\begin{equation}
y_{\text{new}}=\frac{1+\rho-\rho\,y-z\,y^{\rho}}{1-\rho+\rho/y}\;.
\end{equation}
\end{subsection}

\begin{subsection}{Second step: ASEP (from derivatives with respect to $q$)}
\label{appendix numerics Bethe roots: ODEs}
In order to solve the Bethe equations for ASEP, one can consider the $y_{j}$'s as functions of $q$ and take the derivative of the logarithm of Bethe equations (\ref{Bethe equations}) with respect to $q$. We obtain the system of $N$ ordinary differential equations
\begin{equation}
\label{y'[y]}
\fl\hspace{10mm} -\frac{y_{j}'}{1-y_{j}}+\frac{1}{L}\sum_{k=1}^{N}\frac{y_{j}+qy_{j}'-y_{k}'}{qy_{j}-y_{k}}=-\frac{y_{j}+qy_{j}'}{1-qy_{j}}+\frac{1}{L}\sum_{k=1}^{N}\frac{y_{j}'-y_{k}-qy_{k}'}{y_{j}-qy_{k}}\;,
\end{equation}
where $y_{j}'$ is the derivative of $y_{j}$ with respect to $q$. We write $\vec{y}\,'(q)=f(q,\vec{y}(q))$ and evolve (\ref{y'[y]}) from the TASEP solution $\vec{y}(0)$ to a target value $q$ in $M$ steps with increments $\delta q=q/M$ using the modified midpoint method. It consists in writing approximations $Y_{m}$ of $\vec{y}(m\delta q)$, $m=0,\ldots,M$ calculated iteratively by $Y_{0}=\vec{y}(0)$, $Y_{1}=Y_{0}+\delta qf(0,Y_{0})$ and $Y_{m+1}=Y_{m-1}+2\delta qf(m\delta q,Y_{m})$, $m=1,\ldots,M-1$. Then, one chooses as approximation for $\vec{y}(q)$ the quantity $\vec{y}^{M}(q)=\half(Y_{M-1}+Y_{M}+\delta qf(q,Y_{M}))$. The crucial point is that the error from the modified midpoint method has a large $M$, fixed $q$ asymptotic expansion of the form $\vec{y}^{M}(q)-\vec{y}(q)\simeq\sum_{\ell=1}^{\infty}a_{\ell}M^{-2\ell}$ if the integer $M$ is even. Applying the modified midpoint method several times for $M=2,4,\ldots,2K$, it is then possible to use the extrapolation described in section \ref{section numerics} to estimate the value of $\vec{y}(q)$. The combination of the modified midpoint method and rational Richardson extrapolation, usually called the Bulirsch-Stoer method, is a classical method for solving ordinary differential equations with high precision, \textit{i.e.} several hundreds or thousands of digits.
\end{subsection}

\begin{subsection}{Second step: ASEP (from multidimensional Newton's method)}
An alternative method to solve the Bethe equations for ASEP is to use Newton's method iteratively, and increase $q$ slowly enough so that one does not end up with another solution of the Bethe equations. Once one has reached the target value $q$, one can apply Newton's method again with higher precision. Each step of the multidimensional Newton's method used here consists in replacing the $y_{j}$'s with $y_{j}^{\text{new}}$'s solution of the linear system
\begin{eqnarray}
&&\fl\hspace{5mm} \sum_{k=1}^{N}(y_{k}^{\text{new}}-y_{k})\Bigg[\delta_{j,k}\Big(\frac{1}{1-y_{j}}-\frac{q}{1-qy_{j}}\Big)+\frac{\delta_{j,k}}{L}\sum_{\ell=1}^{N}\Big(\frac{1}{y_{j}-qy_{\ell}}-\frac{q}{qy_{j}-y_{\ell}}\Big)\\
&&\fl\hspace{7mm} +\frac{1}{L}\Big(\frac{1}{qy_{j}-y_{k}}-\frac{q}{y_{j}-qy_{k}}\Big)\Bigg]
=\frac{1}{L}\log\Bigg[-\rme^{L\gamma}\Big(\frac{1-y_{j}}{1-qy_{j}}\Big)^{L}\prod_{\ell=1}^{N}\frac{qy_{j}-y_{\ell}}{y_{j}-qy_{\ell}}\Bigg]\;.\nonumber
\end{eqnarray}
This second method, although much cruder than the one of \ref{appendix numerics Bethe roots: ODEs}, performs in practice very well.
\end{subsection}

\end{section}

\end{document}